# Tunable Spike-Timing-Dependent Plasticity in Magnetic Skyrmion Manipulation Chambers

Zulfidin Khodzhaev[1] and Jean Anne C. Incorvia[1]

[1] *Chandra Family Dept. of Electrical and Computer Engineering, The University of Texas at Austin, Austin, TX 78712, USA*

Magnetic skyrmions, as scalable and non-volatile spin textures, can dynamically interact with fields and currents, making them promising for unconventional computing. This paper presents a neuromorphic device based on skyrmion manipulation chambers to implement spike-timing-dependent plasticity (STDP), a mechanism for unsupervised learning in brain-inspired computing. STDP adjusts synaptic weights based on the timing of pre- and post-synaptic spikes. The proposed three-chamber design encodes synaptic weight in the number of skyrmions in the center chamber, with left and right chambers for pre-synaptic and post-synaptic spikes. Micromagnetic simulations demonstrate that the timing between applied currents across the chambers controls the final skyrmion count (weight). The device exhibits adaptability and learning capabilities by manipulating chamber parameters, mimicking Hebbian and dendritic location-based plasticity. The device's ability to maintain state post-write highlights its potential for advancing adaptable neuromorphic devices.

Emulating neural brain functions in solid-state devices is a central focus in neuromorphic computing due to its potential for energy-efficient computing[1,2]. Spike-timing-dependent plasticity (STDP), which adjusts synaptic weight based on the timing of pre-synaptic and post-synaptic spikes[3,4], is fundamental to unsupervised learning in the brain, where synapses independently optimize for learning and memory recall[5]. Fine-tuning the time constant and amplitude of synaptic weight changes is crucial in this process, as it allows the network to adapt the speed and magnitude of synaptic changes to specific computational tasks or environmental conditions. The adaptability of spiking neural networks (SNNs), similar to biological synapses, enhances their learning capabilities, responsiveness, and overall efficiency, proving helpful in managing complex dynamics over extended periods and training neural networks with long-term temporal dependencies[6–8]. Magnetic skyrmions, small magnetic textures found in certain magnetic materials, are attracting attention due to their unique properties[9] with application in computing, particularly their dynamics in confined geometries[10,11], and their potential use as artificial synapses in neuromorphic computing[12].

Previous studies have investigated the behavior of magnetic skyrmions in manipulation chambers and under the influence of temperature[10,11,13], including the design and examination of a skyrmion manipulation chamber for stochastic computing[11]. Moreover, magnetic skyrmions in a magnetic tunnel junction (MTJ) have been shown to act as artificial synapses[12]. However, STDP has not been implemented in multi-skyrmion chamber devices that could be tuned to emulate various forms of synaptic plasticity.

Here, nanodevices based on magnetic skyrmion manipulation chambers are designed to implement STDP. The device is divided into distinct regions, each capable of applying different magnitudes and directions of current. Micromagnetic simulations analyze the device's behavior, demonstrating three different rates of STDP plasticity weight updates and several maximum weight updates. The simulations reveal that the skyrmions' dynamics under various current densities impact the rate of weight updates. This research develops adaptable neuromorphic devices capable of mimicking different types of synaptic plasticity, e.g., Hebbian[3,14] and dendritic location-based plasticity[14].

The STDP model is defined by a function $W(\Delta t)$, where $\Delta t$ represents the difference in timing between post-synaptic and pre-synaptic spikes. The model has different forms depending on the order of neuron spikes[15],

$$W(\Delta t) = \begin{cases} A^+ e^{-\Delta t/\tau^+}, & \Delta t \geq 0 \\ -A^- e^{\Delta t/\tau^-}, & \Delta t < 0 \end{cases}$$

where $\tau^+$ ($\tau^-$) are the time constants for potentiation (depression) that determine the strength of the update over a given interspike interval, $A^+$ and $A^-$ represent the maximum and minimum synaptic changes, respectively, and $\Delta t = t_{post} - t_{pre}$ represents the time difference between the post-synaptic and pre-synaptic spikes.

The design of the device, composed of three chambers, is depicted in Fig. 1(a): two small chambers for the pre-synaptic spike chamber (Pre) and post-synaptic spike chamber (Post), respectively, and a middle weights chamber (W) for weight storage. The synapse weight corresponds to the number of skyrmions in W after a spiking event and can be read using an MTJ[16]. Each chamber has a controllable current density magnitude and direction, managed using contacts $I_1$, $I_2$, $I_3$, $I_4$ and grounds between the middle and small chambers. Two voltage-controlled magnetic anisotropy (VCMA) gates are positioned between the spike and weights chambers, regulated by contacts $V_1$ and $V_2$.

The simulation is executed using MuMax3[17], and the material parameters shown in Table I represent Ir/Co/Pt non-symmetric multilayers[18].

TABLE I. Material parameters used in the simulation.

| Symbol | Magnetic Constant | Values |
| --- | --- | --- |
| $\alpha$ | Gilbert damping | 0.14 |
| $M_s$ | Saturation magnetization | $9.6 \times 10^5$ A/m |
| $K_u$ | Anisotropy constant | $7.17 \times 10^5$ J/m$^3$ |
| $A$ | Exchange stiffness | $1.6 \times 10^{-11}$ J/m |
| DMI | Dzyaloshinskii–Moriya interaction | $1.51 \times 10^{-3}$ J/m$^2$ |

In this study, the term "spike" functions as a self-regulating mechanism similar to the action potential in a biological neuron. The accumulation of skyrmions in the Pre or Post chamber is akin to the electrical potential buildup in a neuron. Once this count reaches 15 skyrmions - the maximum Pre and Post are designed to accumulate before skyrmions start leaving the chamber if gates are turned off - a "spike" is triggered, paralleling a neuron's action potential. This event marks a peak in system activity and causes a reset by turning off applied currents, stopping further skyrmion accumulation. Thus, it serves to regulate system activity and prevent potential overload.

The initial magnetization is uniformly set out of plane ($\hat{z}$) as illustrated in Fig. 1(b). The dimensions of the Pre and Post are 940 nm and 500 nm along the major and minor axis directions, respectively. The W measures 1250 nm and 900 nm in the $\hat{y}$ and $\hat{x}$ directions, respectively. A rectangle of width 600 nm interconnects chambers. The spike chambers have been designed in an elliptical shape to accommodate more stable skyrmions. This results in a smoothed edge between chamber and rectangle regions, preventing edge crashes involving skyrmions[13]. A 5 nm wide strip with a higher anisotropy constant ($K_u$) of $9.27 \times 10^5$ J/m$^3$ covers Pre and Post borders (Fig. 2) to repel skyrmions[19] and improves stability at elevated temperatures. Ref.[13] provides a more detailed examination of temperature influence on manipulation chambers. To expedite the simulations, a damping constant $\alpha = 0.14$ was used, and the simulations were conducted at a temperature of 0 K. The discretization cell is set to $4 \times 4 \times 1$ nm$^3$ in the $\hat{x}$, $\hat{y}$, and $\hat{z}$ directions, respectively.

| $I_{12}$ | $I_{23}$ | $I_{34}$ | $V_1$ | $V_2$ | Function |
|---|---|---|---|---|---|
| $-\hat{y}$ | $-\hat{y}$ | OFF | OFF | OFF | Initial weight |
| $-\hat{y}$ | OFF | OFF | ON | ON | Spike on the left |
| OFF | OFF | $+\hat{y}$ | ON | ON | Spike on the right |
| $-\hat{y}$ | $-\hat{y}$ | OFF | OFF | ON | Long-term potentiation |
| OFF | $-\hat{y}$ | $-\hat{y}$ | ON | OFF | Long-term depression |
| OFF | OFF | OFF | ON | ON | System relaxation |

Figure 2 shows the interplay of currents ($I_{12}$, $I_{23}$, $I_{34}$) and voltages ($V_1$, $V_2$) for the LTP synaptic plasticity function when $\Delta t = 0$, as described in Table II. Details of the process is animated in the supplementary video S1.

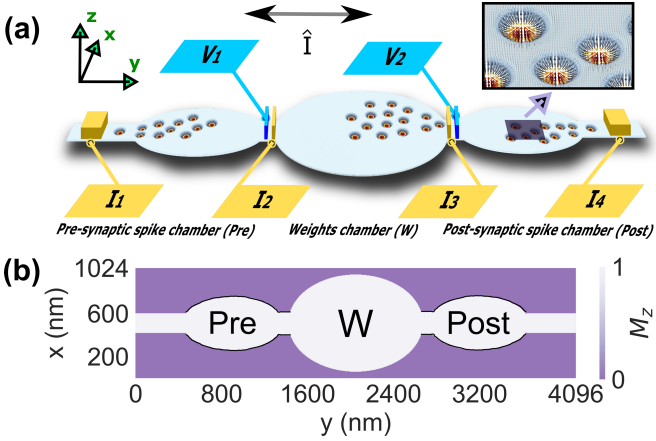

FIG. 1. STDP skyrmion chamber design and simulation setup. (a) The device consists of three chambers (two for spikes, one for weight update) controlled by four contacts ($I_1$-$I_4$) applying current in $\pm\hat{y}$ directions. VCMA contacts ($V_1$, $V_2$) prevent skyrmion overflow. An inset displays simulated skyrmion snapshots. (b) Simulation setup includes initial magnetization and geometric dimensions, with a 5 nm wide strip ($K_u = 9.27 \times 10^5$ J/m$^3$) surrounding Pre and Post regions.

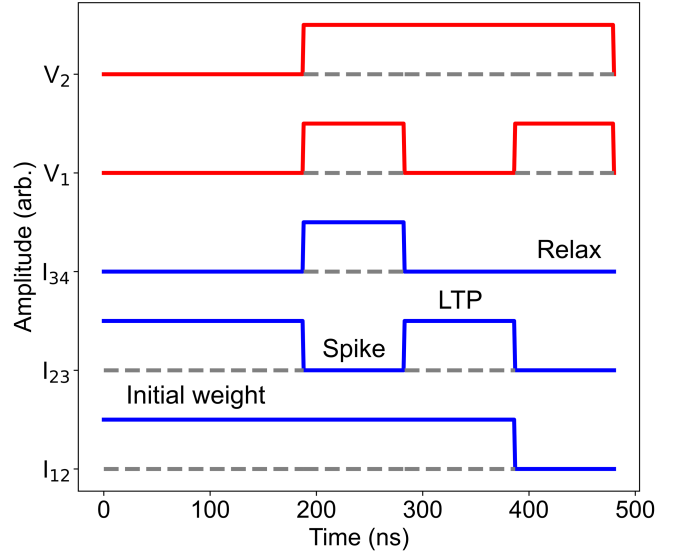

FIG. 2. Sequence and timing of currents ($I_{12}$, $I_{23}$, $I_{34}$), voltages ($V_1$, $V_2$), and key operations (initial weight, spike on the left/right, LTP, system relaxation) in the skyrmion chamber device during the long-term potentiation process.

Table II outlines the range of functions the STDP skyrmion chamber can be tailored to perform, and Fig. 2 shows its timing diagram for the long-term potentiation (LTP) process. The system starts operations with an initial number of skyrmions (initial weight) in the center chamber, which can be introduced via either Pre or Post. In this study, the initial weight was introduced via Pre using $I_{12}$ and $I_{23}$. Next, skyrmions accumulate in Pre and Post using $I_{12}$ and $I_{34}$ currents while $V_1$ and $V_2$ keep skyrmions moving to the W. When the skyrmion count in either chamber reaches a predetermined threshold of 15, $V_1$ is turned off to let skyrmions pass, i.e., the neuron spikes. The VCMA gates increase the anisotropy constant, creating high energy barriers that lead to the annihilation of the skyrmions. The temporal difference between the spikes activates the respective currents for LTP ($I_{12}$, $I_{23}$), and meanwhile, $V_2$ is kept active to prevent overflow from W to Post. The current is applied in the $-\hat{y}$ direction. After the LTP process, the system enters a relaxation stage, where currents are turned off and VCMA gates are kept on until the next spike event is observed.

Following these events, the system undergoes a relaxation phase lasting 100 ns, during which skyrmion-skyrmion repulsion comes into play, increasing the area covered by skyrmions. This demonstrates the device's non-volatility to retain its state post-write. The VCMA gates are simulated using higher anisotropy constant of $K_u = 9.27 \times 10^5$ J/m$^3$ when the gate is on, and $K_u = 7.17 \times 10^5$ J/m$^3$ when the gate is off, returning to the intrinsic anisotropy constant of the material.

TABLE II. Setup for various STDP skyrmion chamber device operations. E.g. $I_{12}$ column indicates the presence and direction of current flowing between electrodes $I_1$ and $I_2$.

Figure 3 shows simulation snapshots and results that illustrate the initial weight transfer and the effect of spike timing delays on the weight update. Figure 3(a-f) plots the initial weight transfer from Pre to W with magnetization snapshots at t = 30 ns (a), 94 ns (b) and 186 ns (c). The arrow indicates the direction of the skyrmion propagation. Figure 3(d) shows the corresponding skyrmion number in the Pre, (e) shows that for the W chamber, and (f) shows for the Post chamber. For initial weight, $I_{12} = 4.5 \times 10^{11}$ A/m$^2$ ($-\hat{y}$) and $I_{23} = 2.5 \times 10^{11}$ A/m$^2$ ($-\hat{y}$) were applied. The ground is connected to both the region between the Pre and W chambers and the region between the Post and W chambers to allow independently supplied currents.

Previous experimental[10] and theoretical[13] work demonstrated that current density is lower in the chamber section of the skyrmion reshuffle device compared to the rectangular sections due to its larger size. This pattern was replicated in modeling the chambers by assigning higher current density at the left rectangle of Pre ($1.8 \times 10^{12}$ A/m$^2$) and lower density within the chamber ($4.5 \times 10^{11}$ A/m$^2$). The same pattern was used for the Post chamber. As a result, skyrmions move quickly towards the spiking chambers and then slow down, allowing time for interaction and accumulation. For the remainder of the paper, only the current density within the spiking chambers will be referred to for both $I_{12}$ and $I_{34}$ currents.

The DBSCAN algorithm[20] from Python's sklearn library was used to detect and count skyrmions in the Pre, W, and Post chambers. Figure 3(g-l) shows the effect of the delay between spike on the left chamber (SL) and spike on the right chamber (SR) on the weight update in W. When the delay is $\Delta t = 0$ ns (Fig. 5(g-i)), both Pre and Post reach 15 skyrmions simultaneously at t = 284 ns, and the weight update is maximized, increasing from 15 to 30 skyrmions in W at t = 401 ns. With a 50 ns delay (Fig. 5(j-l)), the weight increases from 15 to 23 skyrmions, resulting in a lower update. The delay corresponds to the $I_{12}$ and $I_{23}$ currents runtime. For all 15 skyrmions, the runtime to move from Pre to W is 100 ns; for each $\Delta t$, the runtime gets subtracted. VCMA remains active until either Pre or Post reaches the spike threshold. For SL and SR, $I_{12} = 4.5 \times 10^{11}$ A/m$^2$ ($-\hat{y}$) and $I_{34} = 4.5 \times 10^{12}$ A/m$^2$ ($+\hat{y}$) are used. For LTP, $I_{12} = 4.5 \times 10^{11}$ A/m$^2$ ($-\hat{y}$) and $I_{23} = 2.5 \times 10^{11}$ A/m$^2$ ($-\hat{y}$) are applied.

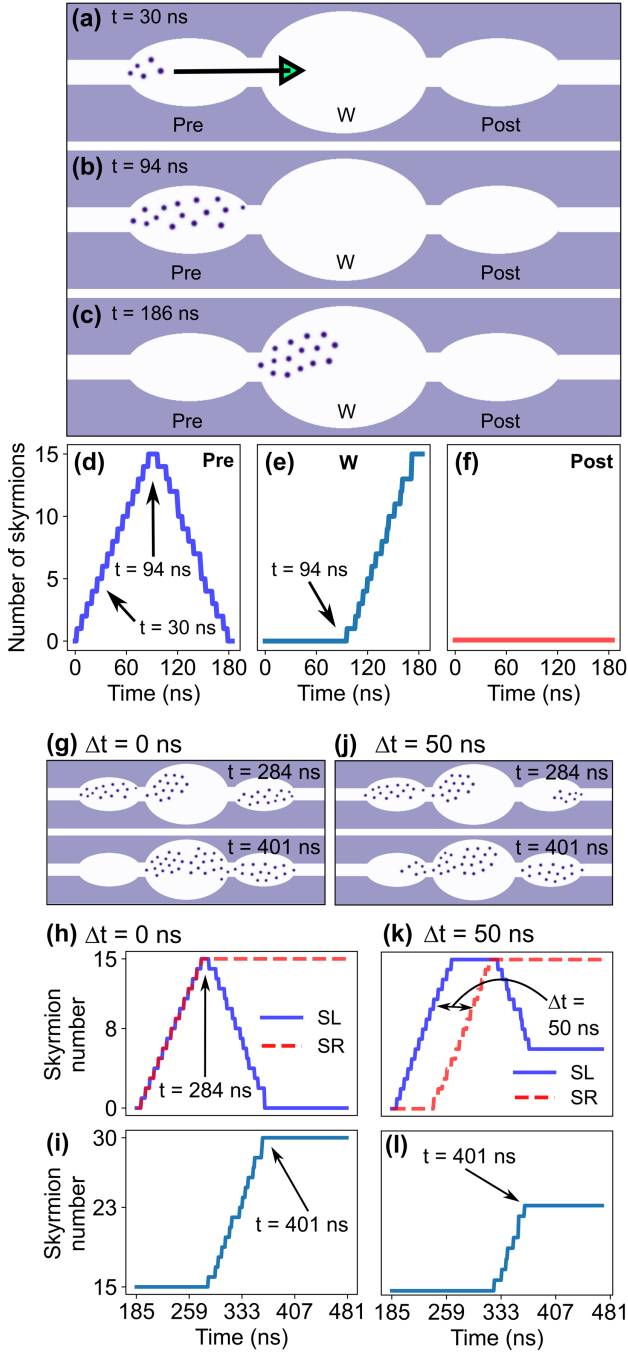

FIG. 3. Initial weight operation and comparative analysis of $\Delta t = 0$ and $\Delta t = 50$ ns spikes for LTP. (a-c) Magnetization simulation snapshots of Pre, W, and Post at 30, 94, and 186 ns. Arrow in (a) indicates skyrmion movement direction. (d-f) Weight updates in Pre, W, and Post. (g-i) Initial weight, skyrmion number over time, and weight update for $\Delta t = 0$. (j-l) Same for $\Delta t = 50$ ns between SL and SR.

Figure 4 shows the LTP and LTD characteristics for different spike timing delays ($\Delta t$) and STDP curves mimicking dendritic location-dependent plasticity. In LTP (Fig. 4(a)), the synaptic weight update inversely depends on $\Delta t$, with the range of changes varying from 0% (15 skyrmions) to 100% (30 skyrmions). The initial jump and drop are attributed to the activation of VCMA gates. In LTD (Fig. 4b), as $|\Delta t|$ increases, the net synaptic weight removed from W decreases, with maximum and minimum changes corresponding to 100% (0 skyrmions) and 0% (15 skyrmions). For LTD, $I_{23} = 3.5 \times 10^{11}$ A/m$^2$ ($-\hat{y}$) ($\Delta t > 0$) and $I_{34} = 4.5 \times 10^{11}$ A/m$^2$ ($-\hat{y}$) ($\Delta t < 0$) are applied. The total runtime is kept at 100 ns by adjusting current densities. Fig. 4(c) shows the progression of STDP from full (100%) to null (0%) synaptic modification, exemplifying Hebbian learning. The results suggest that the skyrmion chamber's weight incorporates the STDP model principles, where synaptic modification depends on the order and relative timing of pre- and post-synaptic spikes, highlighting the complex interplay of factors contributing to synaptic plasticity.

This location dependence arises from neuromodulators like dopamine and acetylcholine, which modulate synaptic plasticity[21–25]. The degree of synaptic strengthening and weakening varies along the dendrite[26]. This principle of STDP-related varying rates of synaptic weight updates is implemented using different current densities within W. Fig. 4(c) illustrates these variations, categorizing STDP rates as strong, moderate, and weak, corresponding to distinct learning locations along the dendrite. For moderate weight, current densities of $I_{12} = 3.5 \times 10^{11}$ A/m$^2$ ($-\hat{y}$) for LTP and $I_{23} = 2.5 \times 10^{11}$ A/m$^2$ ($-\hat{y}$) for LTD were used. For weak weight, $I_{12} = 1.5 \times 10^{11}$ A/m$^2$ ($-\hat{y}$) for LTP and $I_{23} = 0.5 \times 10^{11}$ A/m$^2$ ($-\hat{y}$) for LTD were used.

STDP ranges from 0-100 ns, representing 0 - 100% weight change for a strong rate and 7 - 73% for a moderate rate. The weak rate is 0 - 27% for LTP and 0 - 20% for LTD. Lower skyrmion counts make matching LTP/LTD rates harder for weak updates. However, tunable weight change rates allow for the adjustment of learning behavior, showcasing skyrmion-STDP's versatility.

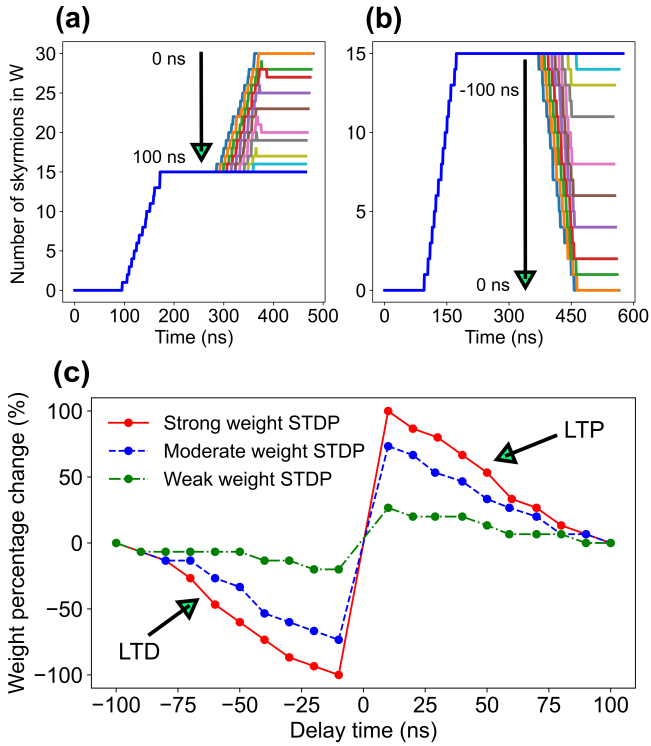

*FIG. 4. STDP plasticity behavior and dendritic location-based plasticity rates. (a) LTP STDP: Skyrmion count in W over time for Δt = 0-100 ns. (b) LTD STDP: Skyrmion count in W for Δt = -100 to 0 ns. (c) Dendritic location-based STDP plasticity rates - strong (0-100% weight change), moderate (7-73%), weak (0-27% LTP, 0-20% LTD).*

The design supports natural skyrmion motion for mimicking plasticity without VCMA gates, as shown in Fig. 5. Material interfacial engineering[27] increases the local perpendicular anisotropy, creating a lower $K_u = 7.38 \times 10^5$ J/m$^3$ in the constriction between chambers. Decreasing the current in Pre to $I_{12} = 1.2 \times 10^{11}$ A/m$^2$ ($-\hat{y}$) achieves natural motion. The Pre chamber holds fewer than 50 skyrmions (Fig. 5(a)), while the weight in W is shown in Fig. 5(b). The mechanism relies on repulsive forces between skyrmions moving through the constriction, as depicted by arrows in Fig. 5(c). Simulation snapshots (Fig. 5(c-e)) illustrate these forces acting on a skyrmion. The decrease in skyrmion diameter signifies the skyrmion overpowering the constriction point energy barriers (Fig. 5(d)) and passing to the other side (Fig. 5(e)). Details are visualized in supplemental video S2. Repulsive forces increase the total energy density within a skyrmion, enabling it to overcome the lowered anisotropy-induced energy barrier. Seventy skyrmions were introduced into Pre, and they move to W until the count reaches 32, beyond which the repulsive effect is insufficient for further movement. Current is only applied to $I_{12}$ and $I_{34}$, while skyrmion-skyrmion repulsion causes expansion and movement inside W. Consequently, plasticity can be implemented using only two contact wires, simplifying the device design.

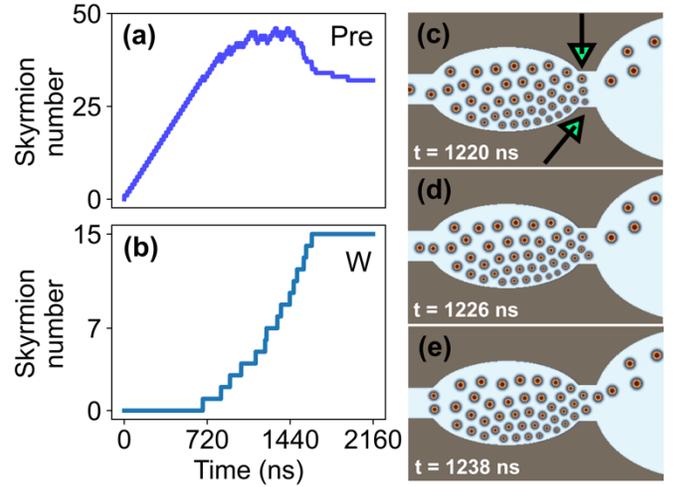

*FIG. 5. Natural motion of skyrmions to mimic plasticity. (a) The weight at Pre shows the 32 skyrmion threshold for leakage to (b) W and (c) shows a magnetization snapshot for skyrmions moving from Pre to W at t = 1220 ns, (d) t = 1226 ns and (e) t = 1238 ns. Arrows show the forces applied to a skyrmion. Color represents $M_z$ from −1 (red) to +1 (blue).*

In summary, these results reveal that magnetic skyrmion chambers can be designed to emulate different types of synaptic plasticity, with STDP plasticity as an example. The chambers offer an adaptable architecture that enables the fine-tuning of STDP behavior by manipulating runtime, applied currents, and applied voltages. Key results include temporal- and diffusion-based control of skyrmion-synapse weight updates, STDP curves mimicking dendritic location-dependent plasticity, and non-volatile synaptic weight storage. These results highlight the potential of skyrmion-based devices to achieve efficient, unsupervised learning in neuromorphic hardware. This biomimetic approach could lead to highly adaptable systems for real-world, online learning applications with further optimization and verification.

This work was supported by the National Science Foundation CAREER under Award Number 1940788. The authors acknowledge the Texas Advanced Computing Center (TACC) at The University of Texas at Austin for providing HPC resources that contributed to the research reported in this paper. URL: http://www.tacc.utexas.edu.

## AUTHOR DECLARATIONS
### Conflict of Interest
The authors have no conflicts to disclose.

## DATA AVAILABILITY
The data supporting this study's findings are available within the article and its supplemental material.